# Medical Image Segmentation with Domain Adaptation: A Survey

Yuemeng Li, Yong Fan

Center for Biomedical Image Computing and Analytics and the Department of Radiology, the Perelman School of Medicine at the University of Pennsylvania, Philadelphia, PA 19104 USA

***Abstract***: Deep learning (DL) has shown remarkable success in various medical imaging data analysis applications. However, it remains challenging for DL models to achieve good generalization, especially when the training and testing datasets are collected at sites with different scanners, due to domain shift caused by differences in data distributions. Domain adaptation has emerged as an effective means to address this challenge by mitigating domain gaps in medical imaging applications. In this review, we specifically focus on domain adaptation approaches for DL-based medical image segmentation. We first present the motivation and background knowledge underlying domain adaptations, then provide a comprehensive review of domain adaptation applications in medical image segmentations, and finally discuss the challenges, limitations, and future research trends in the field to promote the methodology development of domain adaptation in the context of medical image segmentation. Our goal was to provide researchers with up-to-date references on the applications of domain adaptation in medical image segmentation studies.
***Keywords***: medical image segmentation; domain adaptation; deep learning; survey

## 1. Introduction

Deep learning (DL) has been widely applied in the field of medical imaging data analysis studies, including medical image segmentation, registration, detection, classification, and reconstruction to assist clinical diagnosis (Chen et al., 2021e; Du et al., 2021; He et al., 2016; Li and Fan, 2018; Li and Fan, 2020; Li et al., 2023d; Li et al., 2021b; Liu et al., 2021a; Pati et al., 2023; Ronneberger et al., 2015; Zheng et al., 2019). However, DL applications to medical imaging data analysis present unique challenges, particularly with limited labeled images for supervised learning. Although an increasing number of large-scale public datasets have been released to facilitate medical imaging data analysis research (Oakden-Rayner, 2020), DL models are highly sensitive to the training data's distribution. This means that a DL model trained on data from a specific domain might not generalize well to data from a different domain, a problem well known as domain shift (Guan and Liu, 2021). Many studies have explored the transfer learning approach, where a DL model is first trained on a source domain and then fine-tuned on the target domain (Kora et al., 2022). Nevertheless, those methods often demand manual annotations for both domains. Hence, domain shift remains a common challenge for medical image analysis, as variations in imaging techniques, patient demographics, and even device manufacturers can result in vastly different data distributions (Kilim et al., 2022; Kondrateva et al., 2021; Pooch et al., 2020).

Existing review papers have focused on improving the performance of DL models on medical imaging analysis tasks using domain adaptation. In particular, research has shown great potential of generative adversarial networks (GANs) (Goodfellow et al., 2020; Mirza and Osindero, 2014; Zhu et al., 2017) to minimize the domain gap for various medical image applications, such as image synthesis, segmentation, and classification (Kazeminia et al., 2020; Xun et al., 2022; Yi et al., 2019). In addition, Guan *et al.* conducted a review of recent domain adaptation DL models and benchmark medical image datasets for various medical image tasks (Guan and Liu, 2021). Xie *et al.* presented a DL survey on medical image analysis by incorporating domain knowledge for disease diagnosis (Xie et al., 2021). Li *et al.* provided a systematic review for multi-modality cardiac image computing, including domain adaptation approaches (Li et al., 2023b). Despite increasing interest in domain adaptation for medical image analysis, there are limited reviews specifically focusing on medical image segmentation tasks. Therefore, a comprehensive review of DL-based domain adaptation methods for medical image segmentation is desired to identify key challenges and opportunities for future research.

To identify relevant studies for our review of DL methods for domain adaptation-based medical imaging segmentation, we searched databases at PubMed, Semantic Scholar, and Google Scholar with the keywords of “Segmentation”, “Medical Image”, and “Domain Adaptation”. In order to broaden the scope of research and include additional relevant literature, we subsequently checked and included the references and citations of the selected papers from various publishers, including Springer, IEEE Xplore, and Elsevier. As a result, our review provides an extensive summary of 131 peer-reviewed publications with 200 segmentation tasks and 52 publicly available datasets, which were released before or in August 2023, on the topic of domain adaptation-based medical image segmentation.

The remainder of the paper is structured as follows. We first introduced the background knowledge about domain shift, domain adaptation approach, and benchmark datasets in Section 2. We then provide a comprehensive review of medical image segmentation tasks using domain adaptation in Section 3. We categorize all the works based on different body regions. Section 4 discusses the challenges, limitations, and future research trends in medical image segmentations using domain adaptation. Lastly, in Section 5 and Section 6, we present our discussion and conclusion of this review. For tables in Section 3, we use → to represent single direction domain adaptation, and ↔ to represent bi-directional domain adaptation. Furthermore, methods applicable to multiple tasks will be introduced in detail upon their first mention. Subsequent references to the same method, when related to different tasks, can be found in the tables.

# 2. Background

## 2.1. Domain Shift in Medical Image

Domain shift is a common challenge in medical imaging analysis, which is caused by the discrepancies in data distributions, often due to variations in acquisition method, parametrization, and device manufacturing (Kilim et al., 2022; Kondrateva et al., 2021; Li et al., 2022d; Pooch et al., 2020). Particularly, Pooch *et al.* identified the presence of domain shift in chest radiographs across diagnostic centers for lung cancer diagnosis (Pooch et al., 2020). Similarly, Kondrateva *et al.* highlighted the domain shift of MRI data from different acquisition domains in a survey (Kondrateva et al., 2021). Beyond those single domain discrepancies, other studies have discussed the challenges arising from multi-domain shifts, especially when dealing with data variability across distinct image modalities like Computed Tomography (CT) and Magnetic Resonance Imaging (MRI) (Guan and Liu, 2021; Li et al., 2022d). Such domain shifts imply that a model trained on one domain struggles to effectively generalize to another, a problem encountered in medical imaging due to the limited labeled training data. This indicates the importance of minimizing the domain gap between different domains and reducing the dependence on manual labeling.

## 2.2. Domain adaptation Approaches

Transfer learning has been applied to cross-modality medical imaging tasks and has been shown to require less training data compared to traditional DL methods (Kora et al., 2022). Given the training dataset as a source domain and the testing dataset as a target domain, the goal of transfer learning is to transfer the knowledge learned from the source domain to the target domain. This is often achieved by fine-tuning a pre-trained source domain network on data of the target domain. However, the effectiveness of transfer learning is limited by both the availability of labeled data in the target domain and the presence of significant domain gaps between the source and target domains (Guan and Liu, 2021).

To effectively minimize the domain gap cross modalities, various domain adaptation approaches have been proposed. Specifically, generative adversarial networks (GANs) can be utilized for domain adaptation tasks, which learn the potential distribution of data samples and use a generator to produce plausible new data. A discriminator is used to distinguish real data from the generated data by a generator (Goodfellow et al., 2020; Kazeminia et al., 2020). Besides the vanilla GAN (Isola et al., 2017), numerous GAN models have been developed for the domain adaptation-based medical image analysis (Kazeminia et al., 2020; Xun et al., 2022; Yi et al., 2019), such as conditional GAN (cGAN (Mirza and Osindero, 2014)), variational autoencoder GAN (VAE-GAN (Larsen et al., 2016)), and Cycle-GAN (CycleGAN (Zhu et al., 2017)). For instance, Xun *et al.* in the survey demonstrated GAN-based DL methods for medical image segmentation (Xun et al., 2022). Similarly, Kazeminia *et al.* discussed the potential of GAN (Kazeminia et al., 2020).

Other approaches have also been proposed to tackle domain shift problems. The disentangled representation method learns a representation of data that separates the anatomy and modality information, where the anatomy representations can be leveraged for specific tasks (Chartsias et al., 2019). Ensemble learning, which combines the outputs of multiple models to make predictions, has also been used for domain adaptation tasks (Sagi and Rokach, 2018). Knowledge distillation, which effectively distills the knowledge from a larger deep neural network into a small network, has been proposed to transfer knowledge across modalities (Hinton et al., 2015). Meta-learning, aimed at refining a learning algorithm through multiple learning experiences, has been explored in the domain adaptation (Hospedales et al., 2021). Contrastive learning, which enforces the representations to be similar for similar pairs and different for dissimilar pairs, has also shown promising results for cross modality tasks (Chaitanya et al., 2020).

## 2.3. Domain Adaptation-Based Segmentation Categorization

We summarized domain adaptation-based medical image segmentation tasks according to different body regions: head, eye, chest, cardiology, abdomen, pelvis, and others. As illustrated in Figure 1 (a), the cardiology, abdomen, and head areas are the most frequently studied regions. We have further categorized the type of images used for cross-domain segmentation, which include MRI, CT, Funduscopy, Optical Coherence Tomography (OCT), specific types of MRI like Cardiovascular Magnetic Resonance (CMR) and Diffusion-weighted imaging (DWI), and other imaging types such as Xray and colonoscopy. Importantly, when counting each image type in a cross-domain study, we noted every instance. For example, a study involving both MRI and CT was counted under both CT and MRI categories. Figure 1 (b) indicates that MRI and CT are the predominant modalities employed in domain adaptation-based segmentation tasks. Moreover, Figure 1 (c) highlights the growth in research studies focused on domain adaptation-based segmentation tasks since 2018. For a comprehensive review, we summarized all public datasets referenced in Table 1.

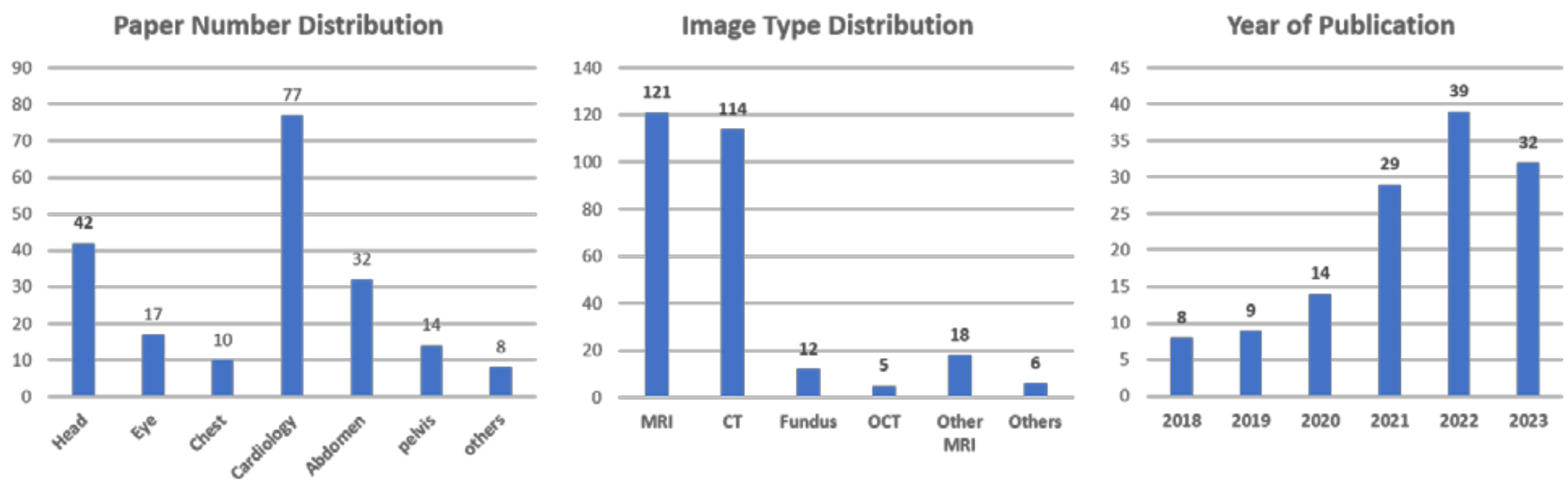

Figure 1. Summary of paper statistics for domain adaptation-based medical image segmentation. (a) Distribution of paper based on segmentation tasks. (b) Distribution of paper based on image types. (c) Distribution of paper based on publication date, ranging from January 2018 to August 2023.

Table 1. A summary of public datasets used for cross-domain medical image segmentation studies. Instance # includes images (2D) and scans (3D) used for training, validation, and testing.

| Task | Dataset | Modality | Instance # |
|---|---|---|---|
| Vestibular Schwannoma and Cochlea | CrossMoDA 2021 (Dorent et al., 2023) | MRI | 210 |
| Vestibular Schwannoma | Vestibular Schwannoma (Coelho et al., 2018) | MRI | 242 |
| Brain Tumor | Brats 2015 (Menze et al., 2014) | MRI | 65 |
| Brain Tumor | ADNI (Trzepacz et al., 2014) | MRI | 147 |
| Brain Lesion | MR Decathlon (Simpson et al., 2019) | MRI | 484 |
| Brain Lesion | CT RSNA Hemorrhage (Flanders et al., 2020) | CT | 15480 2D slices |
| Brain Lesion | MS-SEG (Commowick et al., 2018) | MRI | 15 |
| Brain | iSeg-2017 (Wang et al., 2019a) | MRI | 10 |
| Brain | MRBrainS18[1] | MRI | 7 |
| Brain | HCP (Van Essen et al., 2013) | MRI | 90 |
| Brain | ABIDE (Di Martino et al., 2014) | MRI | 35 |
| Brain | IXI[2] | MRI | 241 |
| Craniomaxillofacial bone | CQ500 (Chilamkurthy et al., 2018) | MRI, CT | 50 |
| Head and Neck | PDDCA[3] | MRI, CT | 48 |
| Spinal Cord Gray Matter | Spinal Cord Gray Matter (Prados et al., 2017) | MRI | 80 |
| Lung tumor | NSCLC[4] | MRI, CT | 377 |
| Spine | IVDM3Seg[5] | MRI | 16 |
| Lung infection | COVID-19 (Ma et al., 2021) | CT | 20 |
| Heart | MMWHS (Zhuang and Shen, 2016) | MRI, CT | 40 |
| Heart | MSC-MRSeg 2019[6] (Zhuang et al., 2022) | MRI CMR | 45 |
| Heart | ACDC (Bernard et al., 2018) | MRI | 200 |
| Heart | RVSC (Petitjean et al., 2015) | MRI | 96 |
| Heart | MyoPS (Zhuang and Li, 2020) | MRI CMR | 225 |
| Heart | Emidec (Lalande et al., 2020) | MRI | 152 |
| Heart | CAMUS (Leclerc et al., 2019) | Echocardiography | 450 |
| Heart | EchoNet-Dynamic (Ouyang et al., 2020) | Echocardiography | 10030 |
| Abdomen | CHAOS (Kavur et al., 2021) | MRI, CT | 80 |
| Abdomen | BTCV[7]/MICCAI 2015 (Landman et al., 2015) | MRI, CT | 47 |
| Abdomen | TCIA-LIHC (Erickson et al., 2016) | MRI, CT | 6 |
| Abdomen | Decathlon[8] | CT | 231 Liver, 40 Spleen |
| Abdomen | WORD (Luo et al., 2022) | CT | 150 |
| Abdomen | NAFLD[9] | CT | 64 |

[1] https://mrbrains18.isi.uu.nl/
[2] http://brain-development.org/ixi-dataset/
[3] https://www.imagenglab.com/newsite/pddca/
[4] https://wiki.cancerimagingarchive.net/display/Public/NSCLC-Radiomics
[5] https://ivdm3seg.weebly.com/
[6] https://zmiclab.github.io/zxh/0/mscmrseg19/
[7] https://www.synapse.org/#!Synapse:syn3193805/wiki/89480
[8] http://medicaldecathlon.com/
[9] https://repository.niddk.nih.gov/studies/nafld_adult/

| | | | |
|---|---|---|---|
| Abdomen | Ircad[10] | CT | 20 |
| Liver | LiTS[11] | MRI, CT | 20 |
| Kidney | CT KiTS19 (Heller et al., 2021) | CT | 210 |
| Knee | Knee Images 2010 (Heimann et al., 2010) | MRI | 60 |
| Prostate | NCI-ISBI13[12] | MRI | 30 |
| Prostate | PROSTATE (Litjens et al., 2012) | MRI | 48 |
| Prostate | I2CVB (Lemaître et al., 2015) | MRI | 19 |
| Prostate | PROMISE12 (Litjens et al., 2014) | MRI | 50 |
| Optic Disc and Optic Cup | REFUGE[13] (Orlando et al., 2020) | Fundus | 800 |
| Optic Disc and Optic Cup | RIM-ONE-r3 (Fumero et al., 2011) | Fundus | 169 |
| Optic Disc and Optic Cup | Drishti-GS (Joshi et al., 2011) | Fundus | 101 |
| Optic Disc and Optic Cup | ORIGA (Zhang et al., 2010) | Fundus | 650 |
| Retina | RETOUCH (Bogunović et al., 2019) | OCT | 48 |
| Retina | OCT (Maunz et al., 2021) | OCT | 44 volumes |
| Retina | UK Biobank (Sudlow et al., 2015) | OCT | 128 |
| Polyp | CVC-Clinic[14] | Colonoscopy | 612 |
| Polyp | ETIS-Larib[15] (Silva et al., 2014) | Colonoscopy | 196 |
| Polyp | EndoScene (Vázquez et al., 2017) | Colonoscopy | 912 |
| Nucleus | KIRC (Irshad et al., 2014) | Histopathology | 486 |
| Nucleus | TNBC (Naylor et al., 2018) | Histopathology | 50 |

# 3. Application of different parts segmentation

## 3.1. Head and neck

Vestibular schwannoma (VS) is a benign tumor that develops on the vestibulocochlear nerve, and cochlea is part of the inner ear for transmitting sound to the brain (Dorent et al., 2023). Studies have shown that one out of a thousand individuals will be diagnosed with VS in their lifetime (Evans et al., 2005). Specifically, C-MADA (Calisto and Lai-Yuen, 2022), Li *et al.* (Li et al., 2022b), COSMOS (Shin et al., 2022), and Dar-UNet (Yao et al., 2022) utilized CycleGAN-based approaches in conjunction with segmentation networks to build segmentation models. While adopting the CycleGAN, Choi *et al.* (Choi, 2022), FPL-UDA (Wu et al., 2022), and MS-MT (Zhao et al., 2023) implemented Contrastive Unpaired Translation (CUT) (Park et al., 2020) that incorporates contrastive learning to the CycleGAN for image translation. BMCAN (Liu et al., 2022b) utilized contrastive learning within a CycleGAN backbone to preserve anatomical details and representations. PAST (Toldo et al., 2020) adopted an extension method of CycleGAN by reusing discriminators for encoding, namely NiceGAN (Chen et al., 2020c), to improve training efficiency. Liu *et al.* (Liu et al., 2022a) utilized CycleGAN followed by label fusion to generate final segmentation. SDC-UDA (Shin et al., 2023) proposed a 2.5D translation framework with both intra- and inter-slice self-attention modules and a pseudo-label refined module for segmentation.

[10] https://www.ircad.fr/research/data-sets/liver-segmentation-3d-ircadb-01/
[11] https://competitions.codalab.org/competitions/17094
[12] https://wiki.cancerimagingarchive.net/pages/viewpage.action?pageId=21267207
[13] https://refuge.grand-challenge.org/
[14] https://polyp.grand-challenge.org/CVCClinicDB/
[15] https://polyp.grand-challenge.org/EtisLarib/

Accurate segmentation of the brain, brain tumors, and lesions is critical for improving clinical diagnosis and treatment planning. 3D cGAN (Yu et al., 2018) utilized a conditional-GAN model with a local adaptive fusion for image synthesize as well as a CNN model for brain tumor segmentation. Furthermore, DSFN (Zou et al., 2020), SASAN (Tomar et al., 2021), JIFAAN (Zhong et al., 2023), SynSeg-Net (Huo et al., 2018b), Dong *et al.* (Dong et al., 2022), and IB-GAN (Tokuoka et al., 2019) implemented CycleGAN in conjunction with segmentation networks. DDA-Net (Bian et al., 2022) incorporated feature adaptations to minimize the feature-level gap across domains with a CycleGAN backbone. Seg-Harmon (Ren et al., 2021) utilized segmentation embedding, while Chen *et al.* (Chen et al., 2019b) proposed a neighbor-based anchoring method, both in a CycleGAN framework. In addition to CycleGAN-based approaches, DSA-Net (Han et al., 2021) proposed a GAN-based approach with a shared encoder and two domain specific decoders to align features, in conjunction with a segmentation model. DLaST (Xie et al., 2022) utilized a GAN-based structure by separating the domain-invariant anatomy from domain specific modality, leveraging disentanglement learning and self-training strategies. Deep-supGAN (Zhao et al., 2018) utilized the GAN framework with deep supervision. Qin *et al.* (Qin et al., 2023) adopted dual student network and adversarial training to tackle domain shift. Liu *et al.* (Liu et al., 2022c) proposed a multi-task framework utilizing edge maps as an auxiliary input to learn contour and guide subsequent segmentation. ACT (Liu et al., 2022d) presented an asymmetric co-training framework to balance the supervision for annotated samples from both source and target domains. Dual-Normalization (Zhou et al., 2022) leveraged the augmented source-similar and source-dissimilar images with a dual-normalization segmentation network. DAE (Karani et al., 2021) proposed a denoising autoencoder for achieving robust image segmentation. PMKL (Chen et al., 2021a) utilized a knowledge learning framework to transfer multimodal knowledge from a teacher network to an unimodal student network with both pixel-level and image-level distillations. OSUDA (Liu et al., 2023d) designed an unsupervised domain shareable batch normalization framework with statistics adaptation. SynthSeg (Billot et al., 2023) utilized a generative model to generate synthetic data during training for segmenting real scans of multi-target domains.

Moreover, other key brain structures have been studied. Chen *et al.* (Chen et al., 2019b) proposed DDA-GAN (Chen et al., 2021f), ARL-GAN (Chen et al., 2020d), and DADASegNet (Chen et al., 2022a) with disentangled learning, representation learning, and attention mechanism respectively for craniomaxillofacial bone segmentation. Perone *et al.* (Perone et al., 2019) utilized self-embedding in a student-teacher model for spinal cord gray matter segmentation. PSIGAN (Jiang et al., 2020) modeled the joint probability distribution via the co-dependency between images for parotid glands in a CycleGAN framework. SC-GAN (Tong et al., 2019) utilized a shape-constraint and a GAN network for head and neck segmentation.

Table 2. Domain adaptation-based image segmentation for brain structures, brain tumor, bone features, and other anatomical regions in the head and neck. List of abbreviations: VSC = Vestibular Schwannoma and Cochlea; VS = Vestibular Schwannoma; CB = Craniomaxillofacial bone; SCGM = Spinal Cord Gray Matter.

| **Publications** | **Code** | **Method** | **Modality** | **Dataset** | **Tasks** |
|---|---|---|---|---|---|
| COSMOS (Shin et al., 2022) | No | CycleGAN | MR ceT1 → hrT2 | CrossMoDA 2021 | VSC |
| PAST (Toldo et al., 2020) | No | NiceGAN | MR ceT1 → hrT2 | CrossMoDA 2021 | VSC |
| Dar-UNet (Yao et al., 2022) | Yes | CycleGAN | MR ceT1 → hrT2 | CrossMoDA 2021 | VSC |
| Li *et al.* (Li et al., 2022b) | No | CycleGAN | MR ceT1 → hrT2 | CrossMoDA 2021 | VSC |
| C-MADA (Calisto and Lai-Yuen, 2022) | No | CycleGAN | MR ceT1 → hrT2 | CrossMoDA 2021 | VSC |
| Liu *et al.* (Liu et al., 2022a) | No | CycleGAN, Label Fusion | MR ceT1 → hrT2 | CrossMoDA 2021 | VSC |
| Choi *et al.* (Choi, 2022) | No | CUT | MR ceT1 → hrT2 | CrossMoDA 2021 | VSC |
| FPL-UDA (Wu et al., 2022) | Yes | CUT | MR ceT1 → hrT2 | CrossMoDA 2021 | VSC |
| MS-MT (Zhao et al., 2023) | No | CUT | MR ceT1 → hrT2 | CrossMoDA 2021 | VSC |
| SDC-UDA (Shin et al., 2023) | No | GAN | MR ceT1 → hrT2 | CrossMoDA 2021 | VSC |
| BMCAN (Liu et al., 2022b) | Yes | CycleGAN | MR T1 ↔ T2 | Vestibular Schwannoma | VS |
| 3D cGAN (Yu et al., 2018) | No | cGAN | MR T1 → FLAIR | Brats | Brain Tumor |
| DSA-Net (Han et al., 2021) | No | GAN | MR T2 → T1, FLAIR, ceT1 | Brats | Brain Tumor |
| IB-GAN (Tokuoka et al., 2019) | No | CycleGAN | MR T2, T1, FLAIR, ceT1 | ANDI, Brats | Brain Tumor |
| SASAN (Tomar et al., 2021) | Yes | CycleGAN | MR T2 → T1 | Brats | Brain Tumor |
| DSFN (Zou et al., 2020) | No | CycleGAN | MR T2 → T1, FLAIR, ceT1 | Brats | Brain Tumor |
| JIFAAN (Zhong et al., 2023) | No | CycleGAN | MR T2 → FLAIR | Brats | Brain Tumor |
| Qin *et al.* (Qin et al., 2023) | No | GAN | MR T2, FLAIR → T1, ceT1 | Brats | Brain Tumor |
| DLaST (Xie et al., 2022) | No | Disentangled Representation | MR T2 ↔ FLAIR | Brats | Brain Tumor |
| Liu *et al.* (Liu et al., 2022c) | No | Semantic Contour | MR T2 → T1, FLAIR, T1c | Brats | Brain Tumor |
| ACT (Liu et al., 2022d) | No | Asymmetric Co-training | MR T2 → T1, FLAIR, T1c | Brats | Brain Tumor |
| Dual-Normalization (Zhou et al., 2022) | Yes | Augmentation and Normalization | MR T2 → T1, FLAIR, T1c; T1c → T1, FLAIR, T2 | Brats | Brain Tumor |
| PMKL (Chen et al., 2021a) | Yes | Knowledge Distillation | MR T2 ↔ T1, FLAIR, T1c | Brats | Brain Tumor |
| OSUDA (Liu et al., 2023d) | No | Batch Normalization | MR T2 → T1, FLAIR, T1c | Brats | Brain Tumor |
| SynSeg-Net (Huo et al., 2018b) | Yes | CycleGAN | CT → MR | Private | Brain |
| Dong *et al.* (Dong et al., 2022) | Yes | CycleGAN | MR → CT | MR Decathlon, CT RSNA Hemorrhage | Brain Lesion |
| BMCAN (Liu et al., 2022b) | Yes | CycleGAN | MR T1 ↔ T2 | iSeg-2017 | Brain |
| DDA-Net (Bian et al., 2022) | No | CycleGAN | MR T1 → T1IR | MRBrainS18 | Brain |
| DAE (Karani et al., 2021) | Yes | Denoising Autoencoder | MR T1 → T1 and T2 | HCP, ABIDE | Brain |
| Seg-Harmon (Ren et al., 2021) | Yes | CycleGAN | MRI | MS-SEG | Brain lesion |
| Seg-Harmon (Ren et al., 2021) | Yes | CycleGAN | MRI | IXI | Brain |
| SynthSeg (Billot et al., 2023) | Yes | GAN | MR T1→ T1, PD, T2, FLAIR, DBS, CT | Multi-center | Brain |
| Chen *et al.* (Chen et al., 2019b) | No | CycleGAN | MR ↔ CT | CQ500, ADNI | CB |
| Deep-supGAN (Zhao et al., 2018) | No | GAN | MR → CT | ADNI | CB |
| DDA-GAN (Chen et al., 2021f) | No | Disentangled Representation | CT → MR | CQ500, ADNI | CB |
| ARL-GAN (Chen et al., 2020d) | No | Representation Learning | CT → MR | CQ500, ADNI | CB |
| DADASegNet (Chen et al., 2022a) | No | Representation Learning | CT → MR | CQ500, ADNI | CB |
| Perone *et al.* (Perone et al., 2019) | Yes | Ensemble Learning | MRI | Spinal Cord Gray Matter Dataset | SCGM |
| PSIGAN (Jiang et al., 2020) | Yes | CycleGAN | CT → MR | Private | Parotid Glands |
| SC-GAN (Tong et al., 2019) | No | GAN | CT ↔ MR | PDDCA | Head and Neck |

## 3.2. Eyes

Accurate segmentation of the optic disc (OD) and optic cup (OC) from color fundus photographs, as well as the retina from OCT, is essential in ophthalmology (Abràmoff et al., 2010; Orlando et al., 2020). As shown in Table 3, DoFE (Wang et al., 2020) explored domain prior knowledge from multi-source domains to enrich image features. SDFA-DPL (Chen et al., 2021b) presented the pixel-level and class-level denoising schemes with uncertainty and prototype estimation for improving pseudo-labeling. DoCR (Hu et al., 2022a) designed an auxiliary high frequency reconstruction strategy. MeFDA (Xu et al., 2021) minimized the entropy maps' prediction of the target domain and used adversarial optimization for the prediction of entropy maps in both domains. DCAC (Hu et al., 2022b) proposed a domain-adaptive module and a content-adaptive module. IOSUDA (Chen and Wang, 2021) introduced image translation to extract shared content features and style features, where shared content features were utilized in a segmentation network. Zhang *et al.* (Zhang et al., 2022) synthesized randomized illumination maps under diverse illumination conditions to improve the segmentation generalization using a retinex-based image decomposition network. ISFA (Lei et al., 2021) utilized content and style feature alignment, along with adversarial learning, for output-level feature alignment using a CycleGAN. FedDG (Liu et al., 2021c) presented a federated learning method to exploit multi-source data distributions. Chen *et al.* (Chen et al., 2021d) utilized an information bottleneck constraint to maintain the consistency of disentangled content features in a CycleGAN architecture. UESM (Bian et al., 2020) proposed an uncertainty-aware cross-entropy loss with a GAN model. SegCLR (Gomariz et al., 2022) proposed a contrastive learning with a contrastive pairing scheme to leverage similarity between nearby slices. RFA (Kang et al., 2023b) introduced a random feature augmentation method to diversify source domain at the feature level without prior knowledges. RDR-Net (Chen et al., 2023b) utilized a dual-path segmentation backbone for edge detection and region prediction with three modules to minimize the domain gap. Chen *et al.* (Chen et al., 2023a) proposed a two-staged CycleGAN network for harmonizing retinal OCT images acquired with different devices. ISTM (Li et al., 2023c) utilized a self-training module, based on a discrepancy and similarity strategy, to select images with better segmentation using a CycleGAN backbone.

Table 3. Domain adaptation-based image segmentation for optic disc and optic cup and retina. List of abbreviations: OD&OC = Optic Disc and Optic Cup.

| **Publications** | **Code** | **Method** | **Modality** | **Dataset** | **Tasks** |
|---|---|---|---|---|---|
| DoFE (Wang et al., 2020) | Yes | Feature Embedding | Fundus | REFUGE (train), REFUGE (val), RIM-ONE-r3, Drishti-GS | OD&OC |
| SDFA-DPL (Chen et al., 2021b) | Yes | Denoising | Fundus | REFUGE (train) → RIM-ONE-r3, Drishti-GS | OD&OC |
| DoCR (Hu et al., 2022a) | Yes | High-Frequency Reconstruction | Fundus | 3 non-public datasets | OD&OC |
| MeFDA (Xu et al., 2021) | No | Encoder-Decoder | Fundus | REFUGE (train) → Drishti-GS, RIM-ONE-r3 | OD&OC |
| DCAC (Hu et al., 2022b) | Yes | Encoder-Decoder | Fundus | All Four | OD&OC |
| IOSUDA (Chen and Wang, 2021) | Yes | Encoder-Decoder | Fundus | REFUGE (train) → Drishti-GS, RIM-ONE-r3, REFUGE (val) | OD&OC |
| Zhang *et al.* (Zhang et al., 2022) | No | Illumination Enhancement | Fundus | REFUGE (train), REFUGE (val), RIM-ONE-r3, Drishti-GS | OD&OC |
| ISFA (Lei et al., 2021) | Yes | CycleGAN | Fundus | REFUGE (train) → Drishti-GS, RIM-ONE-r3 | OD&OC |
| FedDG (Liu et al., 2021c) | Yes | Federated Learning | Fundus | All Four | OD&OC |
| Chen *et al.* (Chen et al., 2021d) | No | Information Bottleneck-GAN | Fundus | REFUGE (train) → REFUGE (val) | OD&OC |
| RFA (Kang et al., 2023b) | No | Feature Alignment | Fundus | REFUGE (train), REFUGE (val), RIM-ONE-r3, Drishti-GS | OD&OC |
| RDR-Net (Chen et al., 2023b) | No | VAE | Fundus | REFUGE (train), RIM-ONE-r3, Drishti-GS, ORIGA | OD&OC |
| Seg-Harmon (Ren et al., 2021) | Yes | CycleGAN | OCT | RETOUCH | Retina |
| UESM (Bian et al., 2020) | No | GAN | OCT | Private | Retina |
| SegCLR (Gomariz et al., 2022) | No | Contrastive Learning | OCT | OCT | Retina |
| Chen *et al.* (Chen et al., 2023a) | No | CycleGAN | OCT | Private, UK Biobank | Retina |
| ISTM (Li et al., 2023c) | No | CycleGAN | OCT | RETOUCH | Retina |

## 3.3. Chest and breast

Accurate segmentation of the lung region, lung infections, and lung tumors is fundamentally important for studies of lung diseases (Hofmanninger et al., 2020). As summarized in Table 4, by adopting the CycleGAN framework, Jiang *et al.* (Jiang et al., 2018) incorporated a tumor-aware loss for lung tumor segmentation, while CMEDL (Jiang et al., 2021) employed knowledge distillation. PSIGAN (Jiang et al., 2020) adopted the CycleGAN backbone for lung tumor segmentation. Chen *et al.* (Chen et al., 2018) utilized the CycleGAN framework for lung segmentation on chest X-rays. CICVAE (Li et al., 2022c) adopted conditional VAE within the CycleGAN model for lung tumor segmentation from MR images. ODADA (Sun et al., 2022b) incorporated a two-step optimization strategy into a GAN model to extract domain invariant representations for lung infection segmentation. DASC (Jin et al., 2021) utilized a self-correction learning strategy for lung infection. MSCDA (Kuang et al., 2023) incorporates self-training with contrastive learning to align feature representations for breast segmentation. UMSA (Zhou et al., 2023) utilized a segmentation network and a GAN network for unpaired lung tumor segmentation.

Table 4. Domain adaptation-based image segmentation for lung region, and breast.

| **Publications** | **Code** | **Method** | **Modality** | **Dataset** | **Tasks** |
|---|---|---|---|---|---|
| Jiang *et al.* (Jiang et al., 2018) | No | CycleGAN | CT → MR | NSCLC | Lung tumor |
| CMEDL (Jiang et al., 2021) | No | CycleGAN | CT → MR | Private | Lung tumor |
| PSIGAN (Jiang et al., 2020) | Yes | CycleGAN | CT → MR | Private | Lung tumor |
| Chen *et al.* (Chen et al., 2018) | No | CycleGAN | Xray Montgomery→ JSRT | Private | Lung |
| CICVAE (Li et al., 2022c) | No | CycleGAN | MR DWI → T2w | Private | Lung Tumor |
| ODADA (Sun et al., 2022b) | Yes | GAN | CT | COVID-19 | Lung infection |
| DASC (Jin et al., 2021) | Yes | Feature Alignment | CT | Private | Lung infection |
| DCAC (Hu et al., 2022b) | Yes | Encoder-Decoder | CT | Private | Lung lesion |
| UMSA (Zhou et al., 2023) | No | GAN | MR DWI → T2w | Private | Lung Tumor |
| MSCDA (Kuang et al., 2023) | Yes | Contrastive Learning | MRI | MR T1 ↔ T2 | Breast |

## 3.4. Cardiology

Accurate whole-heart segmentation is essential for cardiovascular disease studies (Zhuang and Shen, 2016). As shown in Table 5, Segre *et al.* (Segre et al., 2022), CSC-GAN (Zhang et al., 2018), and CoUnetGAN (Cai et al., 2023) utilized the CycleGAN with segmentation models, while other studies implemented CycleGAN with other learning schemes. Particularly, Cui *et al.* (Cui et al., 2021a), SIFA (Chen et al., 2019a; Chen et al., 2020a), and CUDA (Du and Liu, 2021) utilized feature alignment to mitigate the discrepancy between domains. MSGAN (Wang et al., 2022) proposed a multi-stage strategy to transfer detailed texture information on high-resolution feature maps. DALA (Yang et al., 2022b) utilized the attention mechanism to enhance the domain adaptation training. LE-UDA (Zhao et al., 2021; Zhao et al., 2022a) implemented feature adaptations to minimize the feature-level gap across domains. ICMSC (Zeng et al., 2021) utilized semantic consistency to improve image translation. DDFseg (Pei et al., 2021) and CyCMIS (Wang and Zheng, 2022) adopted disentangled representation to improve segmentation. Li *et al.* (Li et al., 2020) proposed a mutual knowledge distillation scheme to exploit the modality shared knowledge. Zhao *et al.* (Zhao et al., 2022b) proposed a transformation consistent meta-hallucination framework to improve image translation. EDRL (Wang et al., 2023b) implemented an entropy-guided disentangled representation learning and a dynamic feature selection mechanism to enhance feature alignment. DEPL (Liu et al., 2023c) utilized domain expansion to learn domain distributions and a selective pseudo-labeling to improve segmentation.

Besides CycleGAN-based approaches, GANs have been widely applied for the domain adaption. PnP-AdaNet (Dou et al., 2019; Dou et al., 2018) and Joyce *et al.* (Joyce et al., 2018) utilized GAN with adversarial training for domain adaptation. PointUDA (Vesal et al., 2021) utilized a point-cloud shape adaptation to tackle domain shift. VarDA (Wu and Zhuang, 2021), PUFT (Dong et al., 2023), and VCEIA (Li et al., 2023e) utilized variational autoencoders to learn latent features from different domains. RSA (Xian et al., 2023) utilized contrastive style augmentation and feature alignment. LMISA (Jafari et al., 2022) proposed a locally normalized gradient magnitude approach with a shape constraint. CDA (Bateson et al., 2021) utilized an inequality constraint to match the prediction statistics from different domains. UMDA-SNA-SFCNN (Liu et al., 2021b) incorporated a spatial attention module into a GAN backbone. Cui *et al.* (Cui et al., 2021b) distilled the extracted 3D landmarks from two domains and incorporated edge information in a segmentation network. ADR (Sun et al., 2022a) leveraged disentanglement learning to extract domain-invariant features via an attention module. SSMNet (Hu et al.) proposed a semantic similarity mining module to improve the domain adaptation. Chen *et al.* (Chen et al., 2020b) implemented a style transfer model and segmentation model followed by fine-tuning on the pretrained segmentation. CFDnet (Wu and Zhuang, 2020) proposed a prior distribution matching to reduce distribution discrepancy between source and target domains. Ge *et al.* (Ge et al., 2023) employed style and boundary information features to facilitate segmentation. Shi *et al.* (Shi and Feng, 2023) applied a Fourier transform-based data augmentation and a self-ensembling model with consistency strategies. Su *et al.* (Su et al., 2023) proposed a global and local feature alignment method to alleviate the domain gap imbalance.

In addition to the GAN-based method, BiRegNet (Ding et al., 2020, 2022) utilized both image registration and label fusion to boost the computational efficiency of the framework. Gong *et al.* (Gong et al., 2021) proposed a conditional domain discriminator and a category centric prototype aligner module to improve the generalizability of the segmentation. MPSCL (Liu et al., 2022e) utilized the contrastive learning with domain adaptative and progressively refined semantic prototypes to exploit the class level distributions. SFDA (Bateson et al., 2022) introduced a source-free domain adaptation framework by minimizing a label-free entropy loss defined in the target domain. UMMKD (Dou et al., 2020) utilized knowledge distillation by constraining the KL-divergence of predicted distributions between modalities. Wang *et al.* (Wang et al., 2023a) proposed a shape-aware alignment loss to align the joint distribution cross-modalities. Bian *et al.* (Bian et al., 2021) utilized a zero-shot learning strategy to extract prior knowledge (relation prototypes) followed by a relation prototype awareness module and an inheritance attention module. ADHC (Chen et al., 2021c) proposed a consistency learning with distribution alignment for left atrium segmentation. DyDA (Jiang et al., 2023) proposed a dynamic credible sample strategy and a hybrid uncertainty learning strategy for multi-domain image segmentation. FSUDA (Liu et al., 2023b) utilized a multi-teacher distillation framework with both frequency and spatial domain transfer. MCTHNet (Liu et al., 2023a) adopted a modality collaborative convolution and transformer hybrid network with representation learning. SEASA (Feng et al., 2023) introduced a data augmentation strategy to generate semantics preserving augmented images followed by semantic alignment for class-level distribution adaptation. Li *et al.* (Li et al., 2023a) utilized a domain translation path to align domain features and a cross-modality segmentation path for segmentation prediction. ADNet++ (Hansen et al., 2023) proposed a feature refined network for multi-class few shot learning.

Table 5. Domain adaptation-based heart segmentation.

| Publications | Code | Method | Modality | Dataset | Tasks |
|---|---|---|---|---|---|
| Cui *et al.* (Cui et al., 2021a) | No | CycleGAN | CT ↔ MR | MMWHS | Whole heart |
| BMCAN (Liu et al., 2022b) | Yes | CycleGAN | MR ↔ CT | MMWHS | Whole heart |
| SASAN (Tomar et al., 2021) | Yes | CycleGAN | MR ↔ CT | MMWHS | Whole heart |
| MSGAN (Wang et al., 2022) | No | CycleGAN | MR ↔ CT | MMWHS | Whole heart |
| DALA (Yang et al., 2022b) | No | CycleGAN | MR ↔ CT | MMWHS | Whole heart |
| DSFN (Zou et al., 2020) | No | CycleGAN | MR → CT | MMWHS | Whole heart |
| LE-UDA (Zhao et al., 2021; Zhao et al., 2022a) | Yes | CycleGAN | MR ↔ CT | MMWHS | Whole heart |
| ICMSC (Zeng et al., 2021) | No | CycleGAN | MR → CT | MMWHS | Whole heart |
| DDFseg (Pei et al., 2021) | Yes | CycleGAN | MR ↔ CT | MMWHS | Whole heart |
| CyCMIS (Wang and Zheng, 2022) | No | CycleGAN | MR ↔ CT | MMWHS | Whole heart |
| SIFA (Chen et al., 2019a; Chen et al., 2020a) | Yes | CycleGAN | MR ↔ CT | MMWHS | Whole heart |
| Segre *et al.* (Segre et al., 2022) | Yes | CycleGAN | MR ↔ CT | MMWHS | Whole heart |
| Li *et al.* (Li et al., 2020) | No | CycleGAN | MR → CT | MMWHS | Whole heart |
| DDA-Net (Bian et al., 2022) | No | CycleGAN | CT → MR | MMWHS | Whole heart |
| CUDA (Du and Liu, 2021) | No | CycleGAN | MR ↔ CT | MMWHS | Whole heart |
| Chen *et al.* (Chen et al., 2021d) | No | Information Bottleneck-GAN | MR → CT | MMWHS | Whole heart |
| Zhao *et al.* (Zhao et al., 2022b) | Yes | CycleGAN | MR → CT | MMWHS | Whole heart |
| DEPL (Liu et al., 2023c) | No | CycleGAN | MR → CT | MMWHS | Whole heart |
| ISTM (Li et al., 2023c) | No | CycleGAN | MR ↔ CT | MMWHS | Whole heart |
| EDRL (Wang et al., 2023b) | No | Disentangled Representation | MR ↔ CT | MMWHS | Whole heart |
| PnP-AdaNet (Dou et al., 2019; Dou et al., 2018) | Yes | GAN | MR → CT | MMWHS | Whole heart |
| Joyce *et al.* (Joyce et al., 2018) | No | GAN | CT ↔ MR | MMWHS | Whole heart |
| DDAGan (Chen et al., 2021f) | No | Disentangled Representation | MR → CT | MMWHS | Whole heart |
| ARL-GAN (Chen et al., 2020d) | No | Representation Learning | MR → CT | MMWHS | Whole heart |
| DADASegNet (Chen et al., 2022a) | No | Representation Learning | MR → CT | MMWHS | Whole heart |
| PointUDA (Vesal et al., 2021) | Yes | GAN | MR → CT | MMWHS | Whole heart |
| VarDA (Wu and Zhuang, 2021) | Yes | VAE-GAN | CT → MR | MMWHS | Whole heart |
| DSA-Net (Han et al., 2021) | No | GAN | MR ↔ CT | MMWHS | Whole heart |
| RSA (Xian et al., 2023) | No | GAN | MR ↔ CT | MMWHS | Whole heart |
| LMISA (Jafari et al., 2022) | Yes | GAN | CT → MR | MMWHS | Whole heart |
| CDA (Bateson et al., 2021) | Yes | GAN | MR → CT | MMWHS | Whole heart |
| ODADA (Sun et al., 2022b) | Yes | GAN | MR ↔ CT | MMWHS | Whole heart |
| UMDA-SNA-SFCNN (Liu et al., 2021b) | No | GAN | MR → CT | MMWHS | Whole heart |
| UESM (Bian et al., 2020) | No | GAN | MR → CT | MMWHS | Whole heart |
| Cui *et al.* (Cui et al., 2021b) | No | GAN | MR → CT | MMWHS | Whole heart |
| Shi *et al.* (Shi and Feng, 2023) | No | GAN | MR ↔ CT | MMWHS | Whole heart |
| DLaST (Xie et al., 2022) | No | Disentangled Representation | MR ↔ CT | MMWHS | Whole heart |
| ADR (Sun et al., 2022a) | Yes | Disentangled Representation | MR ↔ CT | MMWHS | Whole heart |
| SSMNet (Hu et al.) | No | Semantic Learning | MR → CT | MMWHS | Whole heart |
| PUFT (Dong et al., 2023) | No | VAE-GAN | MR ↔ CT | MMWHS | Whole heart |
| BiRegNet (Ding et al., 2020, 2022) | Yes | Label Fusion | MR ↔ CT | MMWHS | Whole heart |
| Gong *et al.* (Gong et al., 2021) | No | Conditional Discriminator | MR → CT | MMWHS | Whole heart |
| Dual-Normalization (Zhou et al., 2022) | Yes | Augmentation and Normalization | MR ↔ CT | MMWHS | Whole heart |
| MPSCL (Liu et al., 2022e) | Yes | Contrastive Learning | MR ↔ CT | MMWHS | Whole heart |
| SFDA (Bateson et al., 2022) | Yes | Class-ratio Loss | MR → CT | MMWHS | Whole heart |
| UMMKD (Dou et al., 2020) | Yes | Knowledge Distillation | MR → CT | MMWHS | Whole heart |
| Wang *et al.* (Wang et al., 2023a) | No | Feature Alignment | MR → CT | MMWHS | Whole heart |
| Bian *et al.* (Bian et al., 2021) | No | Zero-Shot Learning | MR → CT | MMWHS | Whole heart |
| FSUDA (Liu et al., 2023b) | No | Knowledge Distillation | MR → CT | MMWHS | Whole heart |
| MCTHNet (Liu et al., 2023a) | No | Representation Learning | MR ↔ CT | MMWHS | Whole heart |
| SEASA (Feng et al., 2023) | No | Semantic Alignment | MR ↔ CT | MMWHS | Whole heart |
| OSUDA (Liu et al., 2023d) | No | Batch Normalization | MR → CT | MMWHS | Whole heart |
| Ge *et al.* (Ge et al., 2023) | No | GAN | MR ↔ CT | MMWHS | Whole heart |
| SDC-UDA (Shin et al., 2023) | No | GAN | MR → CT | MMWHS | Whole heart |
| Li *et al.* (Li et al., 2023a) | No | Feature Alignment | MR ↔ CT | MMWHS | Whole heart |
| Su et al. (Su et al., 2023) | No | Feature Alignment | MR ↔ CT | MMWHS | Whole heart |
| PointUDA (Vesal et al., 2021) | Yes | GAN | CMR bSSFP ↔ LGE | MSC-MRSeg 2019 | Whole heart |
| Chen *et al.* (Chen et al., 2020b) | No | GAN | CMR bSSFP ↔ LGE | MSC-MRSeg 2019 | Whole heart |
| varDA (Wu and Zhuang, 2021) | Yes | VAE-GAN | CMR bSSFP → LGE | MSC-MRSeg 2019 | Whole heart |
| CFDnet (Wu and Zhuang, 2020) | Yes | GAN | CMR bSSFP → LGE | MSC-MRSeg 2019 | Whole heart |
| RSA (Xian et al., 2023) | No | GAN | CMR bSSFP → LGE | MSC-MRSeg 2019 | Whole heart |
| PUFT (Dong et al., 2023) | No | VAE-GAN | CMR bSSFP → LGE | MSC-MRSeg 2019 | Whole heart |
| DDFseg (Pei et al., 2021) | Yes | CycleGAN | CMR bSSFP → LGE | MSC-MRSeg 2019 | Whole heart |
| CyCMIS (Wang and Zheng, 2022) | No | CycleGAN | CMR bSSFP → LGE | MSC-MRSeg 2019 | Whole heart |
| EDRL (Wang et al., 2023b) | No | Disentangled Representation | CMR bSSFP → LGE | MSC-MRSeg 2019 | Whole heart |
| Li *et al.* (Li et al., 2023a) | No | Feature Alignment | CMR bSSFP → LGE | MSC-MRSeg 2019 | Whole heart |
| VCEIA (Li et al., 2023e) | No | VAE-GAN | CMR bSSFP → LGE | MSC-MRSeg 2019 | Whole heart |
| ADNet++ (Hansen et al., 2023) | No | Few shot | CMR bSSFP → LGE | MSC-MRSeg 2019 | Whole heart |
| CSC-GAN (Zhang et al., 2018) | No | CycleGAN | CT ↔ MR | Private | Whole heart |
| CFDnet (Wu and Zhuang, 2020) | Yes | GAN | CT ↔ MR | Private | Whole heart |
| ADHC (Chen et al., 2021c) | Yes | Consistency Learning | MR → CT | Private | Left Atrium |
| PMKL (Chen et al., 2021a) | Yes | Knowledge Distillation | CMR LGE ↔ T2, bSSFP | Private | Whole heart |
| DAE (Karani et al., 2021) | Yes | Denoising | MRI | ACDC, RVSC | Whole heart |
| DyDA (Jiang et al., 2023) | Yes | Uncertainty Learning | CMR | ACDC, MyoPS, Emidec | Whole heart |
| CoUnetGAN (Cai et al., 2023) | No | CycleGAN | Echocardiography | CAMUS, EchoNet-Daynamic | Heart |

## 3.5. Abdomen

Table 6 summarizes domain adaptation-based segmentation studies for the abdomen and its sub-regions. Particularly, the CycleGAN has been widely applied to this task, such as in Dar-UNet (Yao et al., 2022), SynSeg-Net (Huo et al., 2018a; Huo et al., 2018b), and Sandfort *et al.* (Sandfort et al., 2019). Based on the CycleGAN, MSGAN (Wang et al., 2022) adopted a multi-stage strategy to transfer detailed textures. LE-UDA (Zhao et al., 2022a) implemented feature adaptations to minimize the feature-level discrepancies. SIFA (Chen et al., 2020a) utilized feature alignment to mitigate domain differences. In addition to the CycleGAN-based approaches, GAN-based approaches have been adopted for this task as well. MOvpUNet (Conze et al., 2021) utilized a pretrained generator with a conditional GAN. MASS (Chen et al., 2022b) utilized cross-modal consistency as a constraint to learn generalized domain knowledge. RSA (Xian et al., 2023) utilized a contrastive style augmentation and feature alignment strategy. PUFT (Dong et al., 2023) employed variational autoencoders. Hong *et al.* (Hong et al., 2022a) introduced a joint semantic-aware and shape-entropy-aware adversarial learning for liver segmentation. LMISA (Jafari et al., 2022) utilized a locally normalized gradient magnitude approach for kidney segmentation.

Furthermore, non-GAN approaches have been utilized. SSMNet (Hu et al.) introduced a semantic similarity mining module to improve domain adaptation. BiRegNet (Ding et al., 2022) utilized both image registration and label fusion to boost the computational efficiency. SFUDA (Hong et al., 2022b) combined image and feature alignment with self-training in a knowledge transfer model. LST (Li et al., 2022e) investigated a parameter-free latent space through prior knowledge cross domains. UMMKD (Dou et al., 2020) utilized knowledge distillation. APL (Li et al., 2022a) combined a dual-classifier consistency and predictive category-aware confidence for pseudo-label denoising. Bian *et al.* (Bian et al., 2021) utilized a zero-shot learning strategy followed by a relation prototype awareness module and an inheritance attention module. FSUDA (Liu et al., 2023b) proposed a method based on both frequency and spatial domain transfer under a multi-teacher distillation framework. UDAseg (Li et al., 2021a) employed graph convolutional networks (GCN) and a meta-learning strategy for pancreatic cancer segmentation. DyDA (Jiang et al., 2023) proposed a dynamic credible sample strategy and a hybrid uncertainty learning for kidney and tumor segmentation. OMUDA (Xu et al., 2023) introduced a unified framework for one-to-multiple unsupervised domain segmentation with disentanglement learning.

Table 6. Domain adaptation-based image segmentation for abdomen organs, liver, kidney, and kidney tumor.

| **Publications** | **Code** | **Method** | **Modality** | **Dataset** | **Tasks** |
|---|---|---|---|---|---|
| MSGAN (Wang et al., 2022) | No | CycleGAN | MR ↔ CT | CHAOS, BTCV | Abdomen |
| Dar-UNet (Yao et al., 2022) | Yes | CycleGAN | CT ↔ MR | CHAOS, BTCV | Abdomen |
| LE-UDA (Zhao et al., 2022a) | No | CycleGAN | MR ↔ CT | CHAOS, BTCV | Abdomen |
| SIFA (Chen et al., 2020a) | Yes | CycleGAN | MR ↔ CT | CHAOS, BTCV | Abdomen |
| PSIGAN (Jiang et al., 2020) | Yes | CycleGAN | CT ↔ MR | CHAOS, BTCV, TCIA-LIHC | Abdomen |
| MOvpUNet (Conze et al., 2021) | No | cGAN | CT → MR | CHAOS (Liver) | Abdomen |
| MOvpUNet (Conze et al., 2021) | No | cGAN | MR T1 ↔ MR T2 | CHAOS | Abdomen |
| DSA-Net (Han et al., 2021) | No | GAN | CT → MR | CHAOS, BTCV | Abdomen |
| MASS (Chen et al., 2022b) | Yes | GAN | MR ↔ CT | CHAOS, BTCV | Abdomen |
| RSA (Xian et al., 2023) | No | GAN | MR ↔ CT | CHAOS, BTCV | Abdomen |
| Ge *et al.* (Ge et al., 2023) | No | GAN | MR ↔ CT | CHAOS, BTCV | Abdomen |
| PUFT (Dong et al., 2023) | No | VAE-GAN | MR ↔ CT | CHAOS, BTCV | Abdomen |
| Su et al. (Su et al., 2023) | No | Feature Alignment | MR ↔ CT | CHAOS, BTCV | Abdomen |
| DLaST (Xie et al., 2022) | No | Disentangled Representation | MR ↔ CT | CHAOS, BTCV | Abdomen |
| SSMNet (Hu et al.) | No | Semantic Learning | MR → CT | CHAOS, BTCV | Abdomen |
| BiRegNet (Ding et al., 2022) | Yes | Label Fusion | MR ↔ CT | CHAOS (Liver) | Abdomen |
| Dual-Normalization (Zhou et al., 2022) | Yes | Augmentation and Normalization | MR ↔ CT | CHAOS, BTCV | Abdomen |
| SFUDA (Hong et al., 2022b) | Yes | Knowledge Transfer | CT → MR | CHAOS, BTCV | Abdomen |
| LST (Li et al., 2022e) | No | Latent Space Learning | CT → MR | CHAOS (Liver) | Abdomen |
| UMMKD (Dou et al., 2020) | Yes | Knowledge Distillation | CT → MR | CHAOS, BTCV | Abdomen |
| APL (Li et al., 2022a) | No | Denoising | MR→ CT | CHAOS, BTCV | Abdomen |
| Bian *et al.* (Bian et al., 2021) | No | Zero-Shot Learning | MR → CT | CHAOS, BTCV | Abdomen |
| MCTHNet (Liu et al., 2023a) | No | Representation Learning | MR ↔ CT | CHAOS, BTCV | Abdomen |
| ADNet++ (Hansen et al., 2023) | No | Few shot | MR ↔ CT | CHAOS, BTCV | Abdomen |
| SynSeg-Net (Huo et al., 2018a; Huo et al., 2018b) | Yes | CycleGAN | MR → CT | Private | Abdomen |
| FSUDA (Liu et al., 2023b) | No | Knowledge Distillation | CT → MR | CHAOS, BTCV | Abdomen |
| OMUDA (Xu et al., 2023) | No | Disentangled Representation | CT → MR | BTCV, TCIA, Decathlon, CHAOS, WORD, NAFLD | Abdomen |
| Hong *et al.* (Hong et al., 2022a) | Yes | GAN | CT → MR | CHAOS, LiTS | Liver |
| LMISA (Jafari et al., 2022) | Yes | GAN | CT ↔ MR | CT KiTS19, Private MRI | Kidney |
| LST (Li et al., 2022e) | No | Latent Space Learning | MR → CT | LiTS, CHAOS | Liver |
| Sandfort *et al.* (Sandfort et al., 2019) | No | CycleGAN | Contrast → Non-Contrast CT | Decathlon | Abdomen |
| UDAseg (Li et al., 2021a) | Yes | Meta-Learning | MR T1 ↔ T2 | Private | Pancreatic Cancer |
| DyDA (Jiang et al., 2023) | Yes | Uncertainty Learning | CT | CT KiTS19 and private | Kidney and tumor |

## 3.6. Pelvic

Table 7 summarizes domain adaptation-based image segmentation for pelvic organs. Jia *et al.* (Jia et al., 2019), Brion *et al.* (Brion et al., 2021), and PxCGAN (Kang et al., 2023a) utilized the CycleGAN framework, in conjunction with segmentation models. Lei *et al.* (Lei et al., 2020) utilized the attention mechanism within a CycleGAN backbone for prostate segmentation. ODADA (Sun et al., 2022b) incorporated a two-step optimization strategy to extract domain invariant representations using a GAN model. DAE (Karani et al., 2021) employed a denoising autoencoder for prostate segmentation. DCAC (Hu et al., 2022b) proposed a domain and content adaptive convolution module using an encoder-decoder backbone. SDFA-FSM (Yang et al., 2022a) utilized a Fourier style mining framework to produce high-quality source-like images for prostate segmentation. SAML (Liu et al., 2020) and Wang *et al.* (Wang et al., 2023a) addressed the task of prostate segmentation in MR images through a shape-aware meta-learning scheme and feature alignment respectively. RFA (Kang et al., 2023b) introduced a random feature augmentation method for prostate segmentation from multi-center MRI data.

Table 7. Domain adaptation-based image segmentation for pelvic organs.

| Publications | Code | Method | Modality | Dataset | Tasks |
|---|---|---|---|---|---|
| Lei *et al.* (Lei et al., 2020) | No | CycleGAN | MR → CT | Private | Prostate |
| Jia *et al.* (Jia et al., 2019) | No | CycleGAN | CT → CBCT | Private | Bladder |
| Brion *et al.* (Brion et al., 2021) | Yes | CycleGAN | CT → CBCT | Private | Pelvic Organ |
| ODADA (Sun et al., 2022b) | Yes | GAN | MRI | I2CVB, PROMISE12 | Prostate |
| DAE (Karani et al., 2021) | Yes | Denoising | MRI | NCI-ISBI13, PROMISE12 | Prostate |
| DCAC (Hu et al., 2022b) | Yes | Encoder-Decoder | MRI | Private | Prostate |
| SFDA (Bateson et al., 2022) | Yes | Encoder-Decoder | MRI | NCI-ISBI13 | Prostate |
| SAML (Liu et al., 2020) | Yes | Encoder-Decoder | MRI | I2CVB, PROMISE12 | Prostate |
| Wang *et al.* (Wang et al., 2023a) | No | Feature Alignment | MRI | PROSTATE, PROMISE12 | Prostate |
| SDFA-FSM (Yang et al., 2022a) | Yes | Fourier Style Mining | MRI | NCI-ISBI13, PROMISE12 | Prostate |
| FedDG (Liu et al., 2021c) | Yes | Federated Learning | MRI | I2CVB | Prostate |
| PxCGAN (Kang et al., 2023a) | No | CycleGAN | CT → MRI | Private | Prostate |
| RFA (Kang et al., 2023b) | No | Feature Alignment | MRI | 6 centers | Prostate |

## 3.7. Others

In addition to the above-mentioned segmentation studies, domain adaptation-based segmentation has been applied to various tissues and organs. Specifically, ICMSC (Zeng et al., 2021) leveraged semantic consistency within a CycleGAN backbone for hip joint bone segmentation. SUSAN (Liu, 2019) employed the CycleGAN in conjunction with segmentation models for knee segmentation. MouseGAN++ (Yu et al., 2022) implemented a disentangled and contrastive CycleGAN-based framework for mouse brain segmentation. CDA (Bateson et al., 2021) utilized GAN with an inequality constraint to match the prediction statistics cross domains for spine segmentation, while SFDA (Bateson et al., 2022) introduced a source-free domain adaptation framework. Chen *et al.* (Chen et al., 2021d) utilized information bottleneck constraints for polyp segmentation, while SDFA-FSM (Yang et al., 2022a) utilized a Fourier style mining framework. Yang et al. (Yang et al., 2023) generated pseudo-labels within the CycleGAN framework and fine-tuned a segmentation model based on anatomical assumption for thigh muscle segmentation. NuSegDA (Haq et al., 2023) utilized a segmentation network followed by a reconstruction network and adversarial learning for nuclei segmentation.

Table 8. Domain adaptation-based image segmentation for various tissues and organs, including hip joint bone, knee, spine, polyp, thigh muscle, nuclei, and mouse brain.

| Publications | Code | Method | Modality | Dataset | Tasks |
|---|---|---|---|---|---|
| ICMSC (Zeng et al., 2021) | No | CycleGAN | CT → MR | Private | Hip joint bone |
| SUSAN (Liu, 2019) | No | CycleGAN | MR T1 ↔ T2 | Knee Images 2010 | Knee |
| MouseGAN++ (Yu et al., 2022) | Yes | CycleGAN | MR T1 ↔ T2 | Private | Mouse Brain |
| CDA (Bateson et al., 2021) | Yes | GAN | MR Water → In-Phase | IVDM3Seg | Spine |
| SFDA (Bateson et al., 2022) | Yes | Encoder-Decoder | MR Water → In-Phase | IVDM3Seg | Spine |
| Chen *et al.* (Chen et al., 2021d) | No | Information Bottleneck-GAN | Colonoscopy | CVC-Clinic, ETIS-Larib | Polyp |
| SDFA-FSM (Yang et al., 2022a) | Yes | Fourier Style Mining | Colonoscopy | EndoScene, ETIS-Larib | Polyp |
| Yang et al. (Yang et al., 2023) | Yes | CycleGAN | MR → CT | Private | Thigh Muscle |
| NuSegDA (Haq et al., 2023) | No | GAN | H&E | KIRC, TNBC | Nuclei |

# 4. Open challenges and future directions

## 4.1. Challenges and limitations

### 4.1.1. Limited datasets

Acquiring large-scale medical image datasets that cover a wide range of clinical scenarios remains a significant challenge. Manual annotation of medical images especially on cross-modality datasets is a laborious and expensive task, making it difficult to obtain diverse and large-scale datasets (Krizhevsky et al., 2012) for training domain adaptation models. This difficulty often results in datasets that focus specifically on a single region of interest, as observed in (Dorent et al., 2023; Kavur et al., 2021; Zhuang and Shen, 2016; Zhuang et al., 2022). In clinical practice, however, datasets containing multiple major anatomical structures are essential for tasks like cross-domain disease characterization and surgical planning (Butoi et al., 2023; Wasserthal et al., 2022). The limited availability of large-scale cross-modality datasets and the scarcity of anatomical coverage challenge the development of robust and effective domain models through deep learning.

### 4.1.2. Implementation feasibility

The lack of access to open-source code can also pose challenges for implementing and reproducing results. Factors such as computational resources and deployment feasibility need to be considered when integrating domain adaptation models into clinical environments. Not only is the model performance crucial for clinical research, but gaining recognition from clinicians and patients is also crucial. Providing user-friendly toolkits and straightforward implementation is essential for the adoption of domain adaptation models in the healthcare system (Maleike et al., 2009; Wolf et al., 2005).

### 4.1.3. Multi-modality adaptation

Most DL models discussed in this review were designed for domain adaption studies with single-source and single-target domains and are therefore applicable to just data from one domain. However, applying these models to data from more than two domains presents a challenge. A notable limitation of such domain adaptation approaches is the lack of multi-modality adaptation models for medical image segmentation tasks that cover beyond two data modalities.

## 4.2. Future research trends

### 4.2.1. Transformer model for domain adaptation

In this review, we have discussed DL-based domain adaptation using convolutional neural networks. However, an increasing number of studies demonstrate the effectiveness of transformer architecture with attention mechanisms in medical image analysis (Dosovitskiy et al., 2020; Shamshad et al., 2023; Vaswani et al., 2017). For domain adaptation tasks, feature alignment is commonly employed to extract domain invariant representations (Chen et al., 2019a; Chen et al., 2020a; Cui et al., 2021a; Du and Liu, 2021; Hong et al., 2022b; Lei et al., 2021; Wang et al., 2023a). Meanwhile, other approaches, such as disentangled representation and representation learning, target both domain-invariant and domain-specific feature spaces (Chen et al., 2021d; Pei et al., 2021; Wang and Zheng, 2022; Xie et al., 2022; Yu et al., 2022). This growing interest in efficient transformer architectures could lead to a deeper understanding of multi-modality feature representations for domain adaptation, potentially enhancing network performance in future studies.

### 4.2.2. Source-free domain adaptation

While domain adaptation provides a solution to address domain shift in medical imaging datasets, there remains a need for manual annotations on target datasets for training supervised and semi-supervised domain adaptation (Guan and Liu, 2021). Given the challenge caused by limited manual annotations in medical imaging, recent strategies have introduced unsupervised domain adaptation and zero-shot learning, eliminating the need for labels from target data during model training (Chen et al., 2019a; Chen et al., 2020a; Cui et al., 2021b; Wang et al., 2019b). Such source-free domain adaptation strategies (Fang et al., 2022) reduce the dependency on manual annotations for large-scale datasets. Further exploration of these strategies is highly recommended in the field of medical image segmentation.

### 4.2.3. Multi-task domain adaptation

Future studies are warranted to explore domain adaptations on diverse medical imaging tasks beyond those discussed in this review. This includes expanding beyond specific organs of interest to segment multiple major anatomical structures across modalities. For instance, instead of solely focusing on the heart or abdomen, studies can explore the simultaneous segmentation of both the heart and abdomen within whole-body scans. Such an approach to comprehensive segmentation has significant potential to enhance clinical diagnosis and surgical planning. Additionally, integrating domain adaptation models with existing clinical systems will expand their applicability. Developing user-friendly toolkits (Guan and Liu, 2023; Pati et al., 2023) is crucial for incorporating domain adaptation models into the clinical workflow.

### 4.2.4. Multi-source/ multi-target domain adaptation

Extending domain adaptation beyond two domains is a crucial direction for future research. The studies discussed in the review predominantly focus on adaptations between a single source and target domain. When dealing with multiple domains, these approaches necessitate the training of separate models. This not only leads to increased computational costs but also overlooks the challenges of model generalizability inherent in deep learning medical imaging studies. Given that the majority of existing models are designed for singular target domain tasks, the ability to generalize across multiple domains (Billot et al., 2023; Xu et al., 2023) stands to enhance the adaptability and robustness of these models in diverse clinical environments.

## 5. Discussion

In this review, we have focused on DL-based domain adaptation approaches for medical image segmentation since 2018. We aim to provide a comprehensive overview of the current state-of-the-art DL domain adaptation approaches for cross-modality segmentations in the field of medical imaging analysis. In Section 2, we highlighted domain shifts as a common challenge in cross-modality medical image segmentations and discussed DL-based domain adaptation approaches as an effective solution. We presented literature statistics, including the summary of public datasets, distributions of papers, image modalities, and publication years. It is evident from the analysis that the number of domain adaptation-based approaches has been increasing each year, indicating the growing interest in this area of research.

In Section 3, we categorized the domain adaptation approaches by body regions, including the head and neck, eyes, chest, cardiology, abdomen, pelvis, and others. From the presented approaches, we observed that GANs and their variants, particularly the CycleGAN-based framework, have been widely applied to cross-domain segmentation studies (Xun et al., 2022; Yi et al., 2019). Additionally, other techniques, such as disentangled representation learning, contrastive learning, federated learning, meta-learning, knowledge distillation, and feature alignments, have been investigated to minimize the domain gaps (Guan and Liu, 2021). Furthermore, studies have explored different learning schemes within the CycleGAN framework to enhance model generalizability and achieve robust segmentation (Chen et al., 2019a; Chen et al., 2020a; Cui et al., 2021a; Du and Liu, 2021; Li et al., 2020; Liu et al., 2022b; Pei et al., 2021; Wang and Zheng, 2022).

In Section 4, we discussed the challenges and limitations of implementing domain adaptation models in clinical settings, considering constraints such as limited datasets, implementation feasibility, and multi-modality adaptation. We also highlighted future research trends, emphasizing the exploration of attention-based model architectures, particularly transformers. Furthermore, we underscored the importance of enhancing model accuracy and generalizability through approaches such as source-free, multi-task, and multi-modality domain adaptations for clinical applications and diagnosis.

## 6. Conclusion

We have conducted a comprehensive review of recent DL-based domain adaptation approaches for medical image segmentation tasks. While the reviewed literature demonstrates impressive performance in domain adaptation for medical image segmentation, there are still various unexplored possibilities in this field. We hope that this review will serve as a resource of references to advance and expand the frontiers of domain adaptation research.

## Acknowledgment

This study was supported in part by National Institutes of Health [grant numbers: MH120811, EB022573, AG066650, and DK127488].